\documentclass[twoclumn]{pasj01}

\usepackage[switch]{lineno}
\usepackage{ulem}

\usepackage{comment}

\newcommand{\green}{\textcolor{green}}

\newcommand{\IGR}{{IGR~J00370+6122}}

\usepackage{natbib}

\draft 

\begin{document}
\bibliographystyle{pasj}

\title{A study of the accretion mechanisms of the high mass X-ray binary IGR J00370+6122}
\author{Nagomi Uchida\altaffilmark{1}}%
\author{Hiromitsu Takahashi\altaffilmark{1}}%
\author{Yasushi Fukazawa\altaffilmark{1}}%
\author{Kazuo Makishima\altaffilmark{2,3}}%

\altaffiltext{1}{Department of Physical Science, Hiroshima University, 1-3-1 Kagamiyama, Higashi-Hiroshima, Hiroshima 739-8526, Japan}
\altaffiltext{2}{High Energy Astrophysics Laboratory, RIKEN, 2-1 Hirosawa, Wako, Saitama 351-0198, Japan}
\altaffiltext{3}{Kavli IPMU, The University of Tokyo, 5-1-5 Kashiwanoha, Kashiwa, Chiba 277-8583, Japan}
\email{uchida@astro.hiroshima-u.ac.jp}


\maketitle

\begin{abstract}
IGR J00370+6122 is a high-mass X-ray binary,
of which the primary is a B1 Ib star,
whereas the companion is suggested to be a neutron star 
by the detection of 346-s pulsation in one-off 4-ks observation.
To better understand the nature of the compact companion, 
the present work performs timing and spectral studies of the X-ray data of this object,
taken with XMM-Newton, Swift, Suzaku, RXTE, and INTEGRAL. 
In the XMM-Newton data, a sign of coherent 674 s pulsation was detected,
for which the previous 346-s period may be the 2nd harmonic.
The spectra exhibited the “harder when brighter” trend
in the 1--10 keV range,
and a flat continuum without clear cutoff in the 10--80 keV range.
These properties are both similar to those observed
from several low-luminosity accreting pulsars,
including X Persei in particular.
Thus, the compact object in IGR J00370+6122 is considered to be 
a magnetized neutron star with a rather low luminosity.
The orbital period was refined to $15.6649 \pm 0.0014$ d.
Along the orbit, the luminosity changes by 3 orders of magnitude, 
involving a sudden drop from $\sim 4 \times 10^{33}$ to $\sim 1\times10^{32}$ erg s$^{-1}$
at an orbital phase of 0.3 (and probably vice verse at 0.95).
Although these phenomena cannot be explained by a simple
Hoyle-Lyttleton accretion from the primary's stellar winds,
they can be explained when incorporating the propeller effect
with a strong dipole magnetic field of $\sim 5 \times10^{13}$ G.
Therefore, the neutron star in IGR J00370+6122 
may have a stronger magnetic field
compared to ordinary X-ray pulsars.
\end{abstract}


\section{Introduction}

An important physical parameter of neutron stars (NSs) 
is their magnetic-field (MF) strength $B$, 
because the mechanisms of their accretion and  X-ray emission 
are strongly affected by the values of $B$.
Accurate estimates of the surface MF of accreting NSs 
are available through detections of cyclotron resonance scattering features 
(CRSFs) in their X-ray spectra \citep{Makishima16}.
However, the current observational sensitivity limits the CRSF detections 
to $\lesssim 100$ keV, and hence $B \lesssim 10^{13}$ G.
As an alternative method to estimate the dipole component of $B$ 
(particularly towards higher values) of accreting NSs,
the classical accretion torque theory by Ghosh and Lamb (\citeyear{GL79}), 
hereafter GL79, has been revived, calibrated, 
and applied to several high-mass X-ray binaries (HMXBs)
\citep{Takagi16,Makishima16,Sugizaki17,Yatabe18,Sugizaki20}. 
In particular, \cite{Yatabe18} studied X Persei, 
the HMXB which has an NS companion with 
a long spin period ($\sim$835 s) in an approximate torque equilibrium,
a low luminosity ($\sim$10$^{35}$ erg s$^{-1}$),
and a hard continuum extending to $\sim 80$ keV without a clear cutoff. 
The application of the GL79 modeling has shown 
that the NS in X Persei  has $B \sim$10$^{14}$ G,
which is significantly higher than those of ordinary X-ray pulsars,
and is comparable to those of magnetars.
Such objects, if  plenty, would challenge the consensus 
that magnetars are found solely as isolated NSs.
We are hence urged to search for similar X-ray sources.

According to the results on X Persei,
such an accreting NS with a very strong $B$ is expected to have a long spin period,
and a low luminosity because the implied large magnetosphere will suppress the accretion.
One of the best candidates that fulfill these conditions is \IGR,
the HMXB discovered by INTEGRAL in 2003 (\citealp{denHartog04}).
Its average X-ray intensity measured with the  RXTE/ASM is <1 mCrab and 3 mCrab,
in quiescent and flaring periods, respectively (\citealp{denHartog04}). 
At an estimated distance of 3.4$^{+0.3}_{-0.2}$ kpc (\citealp{Gaia18,Hainich20}),
the latter translates to a 1.5--12.0 keV luminosity of $\sim$10$^{35}$ erg s$^{-1}$.
Furthermore, \cite{Hainich20} analyzed one Swift/XRT observation in quiescence,
and found that the X-ray luminosity is very low at  $\sim 10 ^{32}$ erg s$^{-1}$. 
The optical counterpart is BD+60 73, 
of which the spectral type is either B1 Ib (\citealp{Morgan95}),
B0.5II-III (\citealp{Reig05}), or BN0.7Ib (\citealp{Gonzalez14}). 
The orbital period and eccentricity were measured respectively 
as $15.6627 \pm 0.0042$ d and $0.56 \pm 0.07$ \citep{Zand07, Grunhut14},
or $15.6610 \pm 0.0017$ d and $0.48^{+0.02}_{-0.03}$ \citep{Gonzalez14}.
A circumstellar disk has not been detected (\citealp{Reig05}). 
Thus, \IGR\ indeed has a low luminosity,
even though it must be immersed in dense stellar winds from the early-type companion,
and the binary separation is moderately small.

The compact object in \IGR\ is very likely to be a magnetized NS,
with a suggested spin period of $346 \pm 6$ s, 
because strong flares repeated 7 times with this interval
in an RXTE/PCA observation which lasted about 4 ks \citep{Zand07}. 
The 3-60 keV flare X-ray spectrum, 
which can be modeled by a hard power-law of 
photon index $2.14 \pm 0.02$ \citep{Zand07},
supports this interpretation.
Furthermore, as already suggested by \cite{Grunhut14},
\IGR\ possibly has a rather strong MF, up to $\sim 10^{15}$ G,
because of the relatively long spin period (though still tentative)
and the very low luminosity for its environment.
Another interesting aspect is
that the observed strong flaring activity of this object 
is reminiscent of the behavior of Supergiant Fast X-ray Transients (SFXTs), 
which consist of a supergiant primary and 
a magnetized NS secondary \citep{Bozzo08, Gonzalez14}

Trying to better understand the nature of the compact object in \IGR\
under a working hypothesis 
that it is a NS with a rather high MF,
we analyze, in the present paper, 
the following X-ray data sets of this object. 
First, the 15-year Swift/BAT monitoring data are reanalyzed,
to refine the orbital period, 
and determine accurate orbital phases 
for all the data utilized in this paper. 
Second, we search a 23-ks XMM-Newton light curve, with flaring activity,
for evidence of the pulsation at 346 s or any other periods. 
Then, to characterize the compact object from spectral viewpoints,
we perform spectral analyses (some are orbital-phase resolved)
of the data from XMM-Newton, Suzaku, Swift/XRT, RXTE and INTEGRAL.
Finally, the orbital luminosity modulations are studied,
and are compared  with theoretical calculations 
that assume a simple wind-capture accretion scenario.

\section{Observation and data reduction}
The present work begins with an analysis 
of the orbital intensity modulation of \IGR,
utilizing the Swift/BAT data covering 15 yr from 2005 to 2019.
Then, to derive orbital variations of the X-ray luminosity,
we analyze the data from 34 pointing observations; 
as summarized in in table \ref{table:observation_parameter},
these consist of two observations made with XMM-Newton 
of which one caught a flaring state,
one with the Suzaku/XIS in which the source was quiescent,
and the remaining 31 from the Swift/XRT.
Of these, the flaring-state data set with XMM-Newton 
is employed  also in a search for pulsed signals.
In addition, detailed spectral evaluations are conducted using 
the same XMM-Newton data, those from Suzaku, 
and the 5 brightest of the Swift/XRT data sets.
Finally, to expand the upper energy boundary to $\sim 80$ keV,
we incorporate an RXTE/PCA spectrum taken in 2005 \citep{Zand07},
and a 10-yr average spectrum of \IGR\ taken with  the INTEGRAL/ISGRI
\citep{Walter11}.
All spectral analyses are executed with xspec (version 12.10.1).

\begin{table*}[htb]
\caption{The data sets of \IGR\ utilized in the present paper.}
\begin{center}
 \renewcommand{\arraystretch}{0.85}
  \begin{tabular}{lcccccc}\hline
    Instruments & ObsID & Date (MJD) &\multicolumn{3}{c}{Orbital phase$^{*}$ }\\
                &       &            &  (1)  &  (2)  &  (3) \\
    \hline
    RXTE/PCA
& 91061010101 & 53566.78648-53566.84870 & 0.217-0.221 & 0.120-0.124 & 0.144-0.148\\
    XMM-Newton/EPIC
& 0501450101 & 54505.86816-54506.13731 & 0.173-0.190 & 0.083-0.101 & 0.093-0.110\\
&  0742800201  & 57411.18597-57411.32056 & 0.666-0.674 & 0.596-0.605 & 0.560-0.568\\
    Suzaku/XIS
& 402064010 & 54273.51784-54274.21653 & 0.338-0.383 & 0.247-0.292 & 0.260-0.305\\
    Swift/XRT
&  00032620001 & 56319.00870-56319.27982 & 0.935-0.952 & 0.858-0.875 & 0.838-0.856\\
&  00032620002 & 56327.70678-56327.92077 & 0.490-0.504 & 0.413-0.427 & 0.393-0.407\\
&  00032620003 & 56331.71219-56331.85617 & 0.746-0.755 & 0.669-0.678 & 0.649-0.658\\
&  00032620004 & 56335.57914-56335.66229 & 0.993-0.998 & 0.916-0.921 & 0.896-0.901\\
&  00032620005 & 56339.11291-56339.19714 & 0.218-0.224 & 0.141-0.147 & 0.122-0.127\\
&  00032620006 & 56343.18709-56344.00067 & 0.478-0.530 & 0.402-0.454 & 0.382-0.434\\
&  00032620007 & 56347.25626-56347.33956 & 0.738-0.744 & 0.661-0.667 & 0.641-0.647\\
&  00032620008 & 56351.06961-56351.14633 & 0.982-0.987 & 0.905-0.910 & 0.885-0.890\\
&  00032620009 & 56355.40697-56355.49373 & 0.259-0.264 & 0.182-0.187 & 0.162-0.167\\
&  00032620010 & 56359.47223-56359.75754 & 0.518-0.536 & 0.441-0.460 & 0.421-0.439\\
&  00032620011 & 56363.41796-56363.49504 & 0.770-0.775 & 0.693-0.698 & 0.673-0.678\\
&  00032620012 & 56367.42019-56367.96859 & 0.026-0.061 & 0.949-0.984 & 0.929-0.964\\
&  00032620013 & 56371.09796-56371.37501 & 0.260-0.278 & 0.184-0.201 & 0.163-0.181\\
&  00032620014 & 56379.06295-56379.60814 & 0.769-0.804 & 0.692-0.727 & 0.672-0.707\\
&  00032620015 & 56383.06571-56383.95063 & 0.025-0.081 & 0.948-0.004 & 0.927-0.984\\
&  00032620016 & 56387.40246-56387.95629 & 0.301-0.337 & 0.225-0.260 & 0.204-0.240\\
&  00032620017 & 56391.20790-56391.28544 & 0.544-0.549 & 0.468-0.473 & 0.447-0.452\\
&  00032620018 & 56395.60485-56395.68474 & 0.825-0.830 & 0.749-0.754 & 0.728-0.733\\
&  00032620019 & 56399.80960-56399.89170 & 0.094-0.099 & 0.017-0.022 & 0.996-0.002\\
&  00032620020 & 56403.02594-56403.22481 & 0.299-0.312 & 0.222-0.235 & 0.202-0.214\\
&  00032620021 & 56407.49091-56407.57205 & 0.584-0.589 & 0.508-0.513 & 0.487-0.492\\
&  00032620022 & 56411.36986-56411.56739 & 0.832-0.844 & 0.755-0.768 & 0.734-0.747\\
&  00032620023 & 56416.11359-56416.24773 & 0.135-0.143 & 0.058-0.067 & 0.037-0.046\\
&  00032620024 & 56420.57812-56420.90884 & 0.420-0.441 & 0.343-0.364 & 0.322-0.343\\
&  00032620025 & 57023.68172-57023.75550 & 0.925-0.930 & 0.853-0.858 & 0.822-0.827\\
&  00032620026 & 57024.47586-57024.94858 & 0.976-0.006 & 0.904-0.934 & 0.873-0.903\\
&  00032620027 & 57025.73281-57025.81511 & 0.056-0.062 & 0.984-0.989 & 0.953-0.959\\
&  00032620028 & 57026.67560-57026.74499 & 0.116-0.121 & 0.044-0.049 & 0.014-0.018\\
&  00032620029 & 57027.66365-57027.74351 & 0.180-0.185 & 0.107-0.112 & 0.077-0.082\\
&  00032620030 & 57028.06750-57028.94092 & 0.205-0.261 & 0.133-0.189 & 0.102-0.158\\
&  00032620031 & 57029.19426-57029.94580 & 0.277-0.325 & 0.205-0.253 & 0.174-0.222\\\hline
  \end{tabular}
  \end{center}
(*) Orbital phases are calculated based on the ephemeris obtained by
(1) \citealp{Grunhut14}, (2) \citealp{Gonzalez14},  and (3) in the present work.
\label{table:observation_parameter}
\end{table*}

\subsection{XMM-Newton}
\IGR\ was observed twice by XMM-Newton. 
We reduced these data using the Science Analysis System (SAS) version 18.0,
and generated the cleaned event lists from EPIC PN and EPIC MOS,
utilizing the {\tt emchain} 
and {\tt epchain} tasks, respectively.
In the first observation, the three EPIC detectors (MOS1, MOS2, and PN) 
were all operated in the Small Window mode.
The on-source events with the three EPIC detectors,
covering a 0.1--15 keV range, were extracted 
from a circular region of $35''.0$ radius centred on \IGR. 
Background events of MOS1 and PN were derived 
from a source-free circular region on the same CCD,
with a radius $20''.0$ and $35''.0$, respectively. 
Those of MOS2 were taken from an annular region
with the inner and outer radii of $40''.0$ and $47''.5$, respectively.
In the second observation, the source was observed only in the field-of-view of MOS2 with the Prime Full Window mode.
Source and background photons were collected from a circular region with a radius of $25''.0$ and an annular region with inner and outer radii of $60''.0$ and $70''.0$, respectively.
To produce spectra and light curves,
we use {\tt evselect},
whereas response matrix files (RMFs) 
and ancillary response files (ARFs) are generated
using {\tt rmfgen} 
and {\tt arfgen}, respectively.
The pulse search is carried out using the XRONOS software package, 
distributed by High Energy Astrophysics Science Archive Research Center (HEASARC).
For that purpose,
we applied the barycentric corrections to all the photons,
by {\tt barycen} in the SAS software.

\begin{figure*}
\centering
  \includegraphics[width=170mm, angle=0]{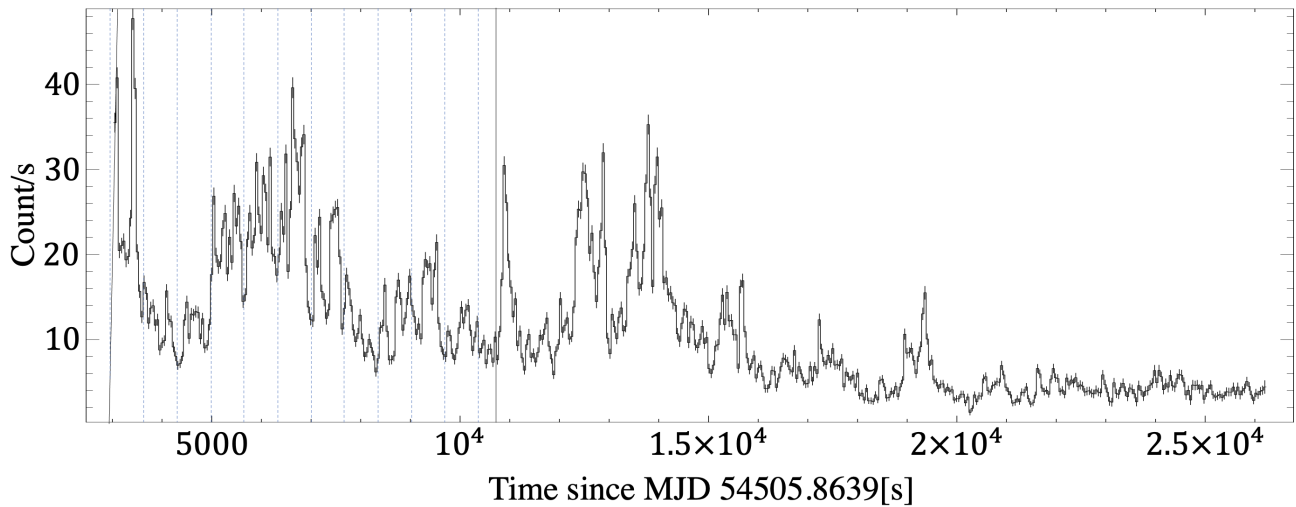}
 \caption{Background subtracted 0.1--15 keV light curve of \IGR,
acquired in the first XMM-Newton observation with EPIC/PN.
The vertical lines indicate the 674 s periodicity.}\label{fig:xmm-newton_lc}
\end{figure*}

In the first observation, 
made  in 2008 for a net exposure of 23 ks,
the source exhibited a strong flaring activity
around a mean luminosity of $\sim 1 \times 10^{35}$ erg s$^{-1}$.
In the second XMM-Newton observation conducted in 2016 
for a net exposure of 11 ks,
the source luminosity was much lower, 
$\sim 1 \times 10^{32}$ erg s$^{-1}$.
Figure \ref{fig:xmm-newton_lc} shows 
an EPIC/PN light curve from the 1st observation, 
in the 0.1--15 keV energy range
(without energy selection in the events).
In the first half of the observation,
we observe high count rates and strong intensity variations.
The MOS1 and MOS2 data acquired in this observation
are incorporated later in the spectral analyses.

\subsection{Suzaku/XIS}
The archival Suzaku data of \IGR, in the 0.2--12 keV range,
were acquired in 2007 (MJD 54274).
Of the four XIS cameras, we utilize the data from XIS0, XIS1, and XIS3, 
which were operated with 1/4 window mode at that time. 
The HXD data are not employed, because \IGR\ was not detected by the HXD.
We processed and screened the XIS data with 
{\tt aepipeline} 1.1.0. 
and extracted source and background event with {\tt xselect}.
The source events were accumulated  over a circular region 
with $110''$ radius around the source position.
The background events were derived from two rectangular regions,
both having a size of $ 180''\times 216''$ 
and located at  $0^\circ .1$ off the source.
The RMFs and ARFs were generated using {\tt xisrmfgen} 
and {\tt xisarfgen} in HEASoft, respectively.

\subsection{Swift/BAT and Swift/XRT}
To improve the orbital ephemeris of \IGR,
we use its 15--50 keV Swift/BAT daily light curve
from 2005 February to 2019 November,
provided by the Swift/BAT Hard X-ray Transient Monitor project (\citealp{Krimm13}).

\IGR\ was observed also with the Swift/XRT,
24 times during  $\sim 100$ days in 2013 (MJD 56319-56420),
and 7 days in 2015 (MJD 57023-57029).
These photon-counting-mode data sets, 31 altogether, 
are also analyzed in the present work.
We processed each data set, consisting of 0.2--10.0 keV photons,
using {\tt xrtpipeline}
and produced the spectra and light curves 
using {\tt xrtproducts}.
The on-source events were extracted from  a circle of $47''$ radius,
and the background events from an annulus
with the inner and outer radii of $94''$ and $141''$, respectively.
\footnote{.
The original light curve downloaded from the web page (\citealp{Krimm13})
had the fits header keyword of TIMEPIXR=0.5,
meaning that the TIME value corresponds to the middle of the bin.
We modified it from 0.5 to 0,
so that the TIME value becomes the beginning of the bin 
(private communication with the Swift/BAT team at NASA/GSFC).}

\subsection{The other data sets}
The RXTE/PCA archival data, 
in which \cite{Zand07} detected the 346-s pulsation, 
were used to characterize the spectrum of \IGR\
in an intermediate energy range (7--30 keV).
The background and response files are included in the same archive.
We ignored <8 keV region to avoid Xenon L-edge feature.

The reduced INTEGRAL/ISGRI spectrum was downloaded 
from the site of High-Energy Astrophysics Virtually ENlightened Sky (HEAVENS),
provided by the INTEGRAL Data Centre (ISDC). 
We selected 17.3--80.0 keV good-quality events
that were acquired over 2003--2012 
and in an orbital phase interval from -0.1 to 0.1.
The background and response matrix were generated also by HEAVENS.

\section{Data Analysis and Results}
\subsection{Orbital intensity variations}

\begin{figure}
\centering
  \includegraphics[width=120mm, angle=0]{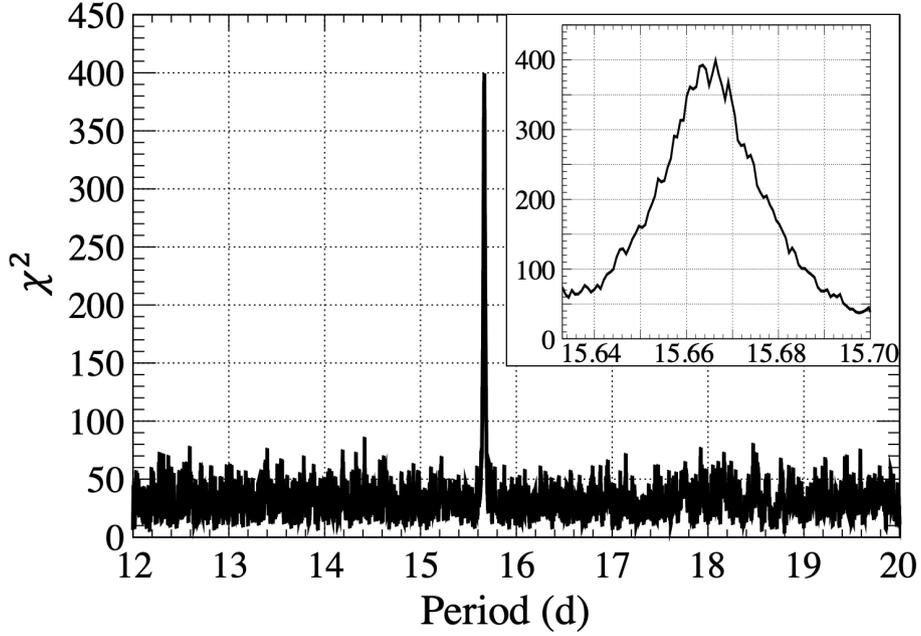}
 \caption{Periodgram of the 15-yr Swift/BAT data,
 revealing the orbital period.
 The inset shows an enlargement around the 15.7-d peak.}
 \label{fig:swift_bat_chisquare_dist}
\end{figure}

As the first attempt of our data analysis,
the 15-yr Swift/BAT light curve in 15--50 keV was analyzed
for the expected 15.7 d orbital periodicity \citep{Zand07,Gonzalez14},
employing the standard epoch-folding method with chi-square evaluation.
We used the {\tt efsearch} software with 32 phase bins,
and scanned the trial period from 12 d to 20 d with a step of 0.0007 d.
The obtained  periodogram is shown in figure \ref{fig:swift_bat_chisquare_dist}. 
A strong peak seen around 15.66 d represents the orbital period,
as reported in the previous works. 
By fitted this peak with a Gaussian,
we determined the orbital period as 
\begin{equation}
P_{\rm orb}= 15.6649 \pm 0.0014~{\rm d}
\label{eq:Porb}
\end{equation}
where the error refers to 90\% confidence level.
This $P_{\rm orb}$ is consistent with 
the previous measurements \citep{Zand07,Gonzalez14}, 
and has a better statistical accuracy,
because we used a longer observation span.

\begin{figure}
\centering
 \includegraphics[width=120mm, angle=0]{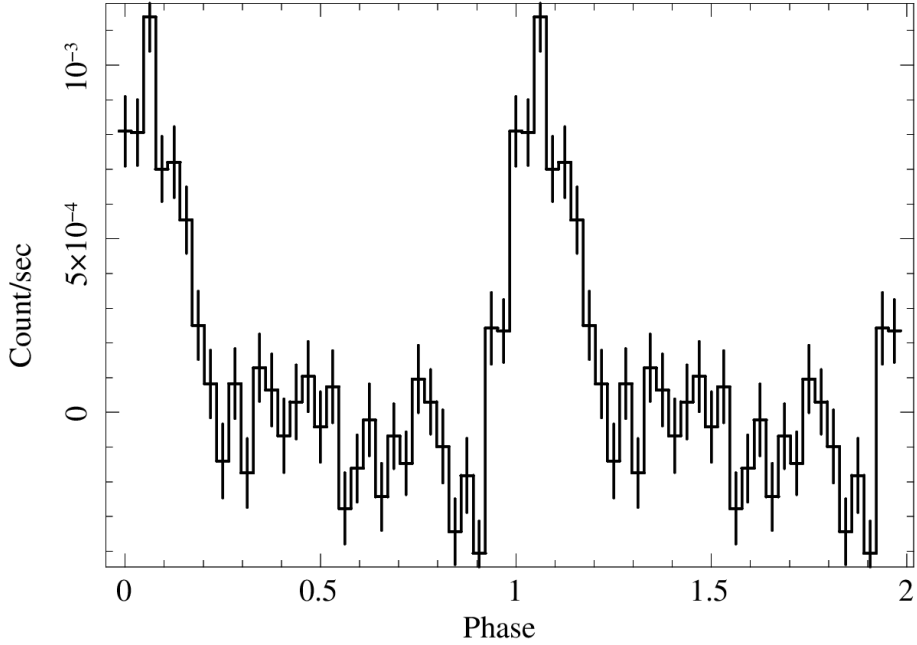} 
 \caption{The Swift/BAT data, folded at $P_{\rm orb}$ of 
 equation~(\ref{eq:Porb}) and shown for two cycles.
 The epoch of phase zero is set as the periastron time given in \cite{Gonzalez14}.}
 \label{fig:swift_bat_orbital_folded_lc}
\end{figure}

Figure \ref{fig:swift_bat_orbital_folded_lc} is 
the 15--50 keV Swift/BAT light curve folded at $P_{\rm orb}$,
where phase 0 is taken as the periastron time of 
\begin{equation}
    {\rm MJD} \: 55084.018
    \label{eq:periastron} 
\end{equation}
as determined by \cite{Gonzalez14}.
Thus, the source exhibits a strong orbital intensity variation,
and has been detected with the Swift/BAT
over a limited orbital phase of about $-0.1$ to 0.25.
In addition, the intensity maximum is clearly delayed
from the periastron by $\sim 0.05$ orbital cycle,
although the delay could be slightly smaller 
than the previously reported $0.1-0.2$ cycles \citep{Grunhut14,Gonzalez14}.

In table 1, we summarize the orbital phases of the 
individual observations analyzed in the present work,
where the three columns for the orbital phase,
(1) to (3), employ the three orbital solutions;
\cite{Grunhut14}, \cite{Gonzalez14}, and the present work.
Below, we utilize our own phase values,
which are based on  $P_{\rm orb}$ of equation~(\ref{eq:Porb})
and the phase 0 (periastron) epoch  of equation~(\ref{eq:periastron}).
\subsection{Search for pulsations}
One of the most important objectives of the present study is to confirm
that the compact object in \IGR\ is a magnetized NS
powered via mass accretion.
Evidently, the best way towards this goal is 
to detect a periodic source pulsation,
at a period that is consistent with, 
or closely related to, the previously recorded 346-s periodicity.
For this purpose, we choose the 1st observation with XMM-Newton,
because the source was bright at that time,
and exhibited multiple flares (figure \ref{fig:xmm-newton_lc})
which are reminiscent of the case with  \cite{Zand07}.

\subsubsection{Fourier power spectra}
Using the background-subtracted 0.1--15.0 keV EPIC/PN light curve 
shown in figure \ref{fig:xmm-newton_lc} from this observation,
and employing {\tt powspec} with a time resolution of 1 s 
with which we generated the light curve,
we calculated a Fourier power spectrum over a frequency range of 
$4\times 10^{-5}-0.5$ Hz.
The result is presented in 
figure~\ref{fig:xmm-newton_pulse_search_fft} in black,
where the abscissa (i.e., the frequency) is
shown in a logarithmic scale 
to better reveal the low-frequency region.
The power is normalized so as to become 2.0
when the light curve is dominated by the Poisson white noise.
Although a strong red noise component emerges at $\lesssim 1$ mHz,
we do not find any preferred periodicity in the analyzed range,
including in particular at $\sim 2.9$ mHz
which corresponds to 346 s.

As seen in figure \ref{fig:xmm-newton_lc},
the source repeated spiky flares in the 1st half of the observation,
and became much quiescent in the 2nd half.
Therefore, we next calculated the Fourier power spectra 
for different time regions of this light curve.
Red and blue traces in the same figure are 
the results for the first 8 ks 
(i.e., 3--11 ks in figure \ref{fig:xmm-newton_lc}) 
and the remaining part of the light curve, respectively.
The 1--10 mHz frequency regions of these spectra
are expanded in the inset to figure
\ref{fig:xmm-newton_pulse_search_fft}(left).
There, the power is still two orders of magnitude higher 
than would be expected for a pure Poissonian noise,
implying that the red noise extends into these frequencies.
Superposed on the red noise, the spectrum from the 1st 8 ks
exhibits several noticeable peaks,
including the highest one at 1.5 mHz,
and the 2nd highest one at 2.9 mHz.
As evident from the logarithmic top abscissa of the inset,
they are apparently in the 1:2 harmonic ratio,
and the 4th harmonic could also be seen at 5.6 mHz.
This suggests the presence of a 1.5 mHz periodicity
(together with its higher harmonics) in the 1st 8 ks of the data,
and the 2.9 mHz feature (possibly the 2nd harmonic)
could be identified with the 346-s periodicity.

It is however not easy to
tell whether these peaks are real,
or just due to fluctuations in the strong red noise.
We hence adopt the technique by Israel and Stella (\citeyear{Israel96}), 
which allows us to evaluate the significance 
of a suggested periodicity in a power spectrum
where strong colored noise is present.
It compares the power at a particular target frequency,
with the local continuum estimated by averaging the spectrum
on both (high-frequency and low-frequency) 
sides with logarithmically symmetric widths.
Employing this technique with the smoothing width of 30 wave numbers,
we obtained the results 
shown in the right panel of Figure \ref{fig:xmm-newton_pulse_search_fft}. 
The smoothed power spectrum, plotted in blue,
roughly follows a power-law function with an exponent of -1.5, which is typical of red noise.
The red data points (with linear ordinate), obtained
by dividing the raw power spectrum by the smoothed one, 
are confirmed to closely follow a
$\chi^2$ distribution with 2 degrees of freedom.

There, we observe a series of prominent peaks,
at 1.47, 2.93, 5.62, 9.03, and 11.72 mHz.
Within the frequency resolution determined by the data length,
their ratios,
$1.00:1.99:3.82: 6.14: 7.97$,
are consistent with the (mostly even) harmonic ratios 1:2:4:6:8.
Among them, the one at 1.47 mHz,
to be identified with the fundamental,
is obviously the feature noticed in the left panel,
and has a chance-occurrence probability of ${\cal P}_1=0.0198$
before correcting for the frequency trials.
The highest one at 11.72 mHz, regarded as the 8th harmonic,
has a normalized power of 11.7,
with the pre-trial probability of ${\cal P}_8 = 3.8 \times 10^{-3}$.

Let us suppose that the 1.47 mHz peak appeared
due to a chance fluctuation, with a probability ${\cal P}_1$.
Then, the probability to observe also the highest peak,
at a frequency corresponding just to the 8th harmonic,
should be ${\cal P}_1 {\cal P}_8=7.5 \times 10^{-5}$.
Further multiplying the frequency trial number, 121,
we finally obtain the overall false alarm probability of $0.9\%$.
Considering however the ambiguity in choosing the smoothing width,
and uncertainties in treating the harmonics,
we conservatively quote a probability of a few percent.
We therefore conclude, with a confidence of $>90\%$,
that the source showed evidence of 1.47 mHz periodicity,
or pulsation with a period of $\sim 680$ s,
at least during the 1st 8 ks of the data.
Incidentally, the chance probability would further decrease
if we consider the simultaneous presence 
of the 2nd, 4th, and 6th harmonics,
but this would be an overuse of the harmonic condition.

\subsubsection{Periodogram analysis}
\begin{figure*}
\centering
  \includegraphics[width=170mm, angle=0]{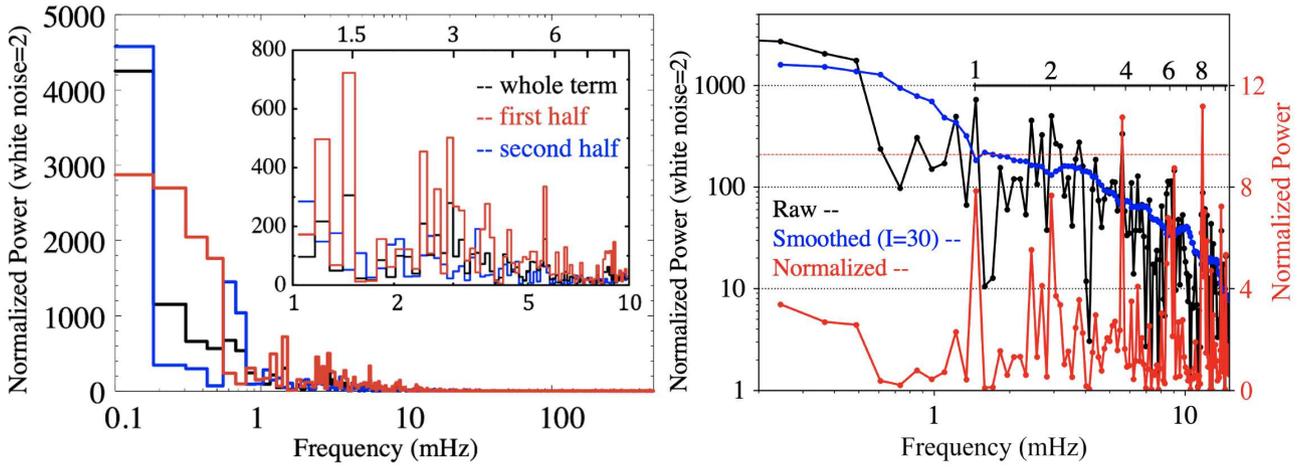}
 \caption{
(Left) Fourier power spectra 
for the whole (black), the first 8 ks (red), 
and the remaining part (blue) of the XMM-Newton/PN light curve.
The abscissa is logarithmic, in order to reveal low-frequency structures.
The inset is the enlarged view of the spectra in the 1--10 mHz range,
where the logarithmic tick marks at the top abscissa are drawn 
at multiples of 1.5 mHz.
(Right) Evaluation of the power spectrum from the first 8 ks,
using the technique by Israel and Stella (\citeyear{Israel96}). 
The raw spectrum in black (the same as red in the left panel)
is averaged over 30 data points,
into a smoothed spectrum shown in blue.
Both use logarithmic ordinate.
The red solid line presents the normalized spectrum,
where the raw spectrum is divided by the smoothed one.
It refers to the linear ordinate on the right,
and the inset abscissa shows the harmonic numbers.
The horizontal dashed line in red indicates 
the 99\% confidence level for a single trial.
}
 \label{fig:xmm-newton_pulse_search_fft}
\end{figure*}

\begin{figure*}
\centering
  \includegraphics[width=170mm, angle=0]{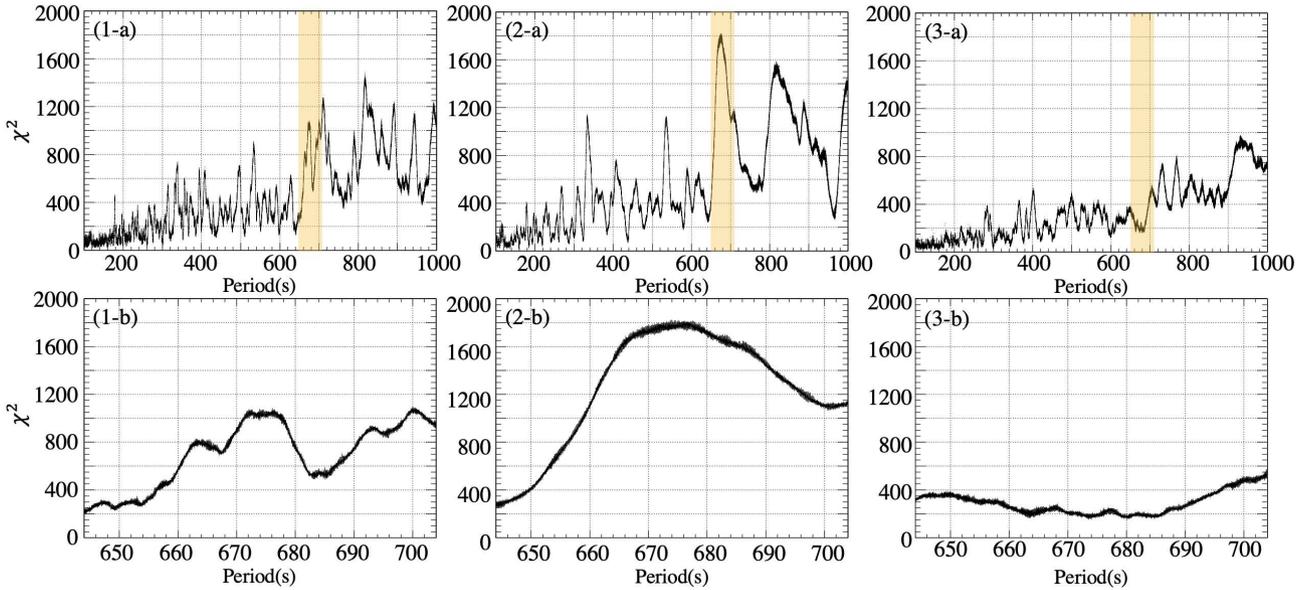}
 \caption{Chi-square periodograms with 32 bins,
 derived from (1-a) the whole data, 
 (2-a) the first 8 ks,  and (3-a) the latter 15 ks.
 Panels (1-b) to (3-b) expands these periodograms  
 over a period range of 645--705 s. 
 }
 \label{fig:xmm-newton_pulse_search_periodogram}
\end{figure*}

Although we have obtained evidence for the 1.5 mHz periodicity,
the limited frequency resolution
does not allow an accurate period determination.
Thus, following a standard procedure in pulsar studies,
we next calculated chi-square periodograms using {\tt efsearch} with 32 bins, over the frequency range of 1--10 mHz
corresponding to the inset to figure \ref{fig:xmm-newton_pulse_search_fft},
or equivalently, a period range of 100--1000 s,
with a step of 0.1 s.
The obtained results are shown 
in figure~\ref{fig:xmm-newton_pulse_search_periodogram},
where panels (1-a), (2-a), and (3-a) represent
the entire data, the 1st 8 ks, and the latter 15 ks,
respectively.

In agreement with the power spectra with the strong red noise,
the obtained chi-square values are much larger than the degree of freedom,
essentially at any period studied here.
Furthermore, in panel (2-a) derived from the 1st 8 ks of the data,
we observe several prominent peaks,
although they are not recognized 
when we use the entire 23-ks data (panel 1), 
or those from the latter 15 ks (panel 3).
The highest peak at $\sim 675$ s,
of which the details are shown in panel (2-b),
obviously corresponds to the 1.47 mHz periodicity
found with the power spectrum.
By fitting this peak with a cubic function,
we determined the best period, 
or the maximum point of the probability distribution, as 674 s.
A Gaussian fitting has given the same result.
Although it is not straightforward to define the associated error,
we conservatively quote a value of $\pm 15$ s,
as the half-width at half-maximum of the chi-square peak above the background.

In order to cross-check the result obtained
using the normalized power spectrum,
the significance evaluation was conducted 
also using the chi-square periodogram.
For this purpose, we Fourier synthesized 1000 fake light curves,
in which each Fourier amplitude is set 
to the square root of the normalized power 
obtained in the previous subsubsection, 
whereas the associated Fourier phase is randomized.
By analysing the fake light curves in the same way as the actual data,
we studied the probability of finding chi-square values,
exceeding what was observed (about 1800), at the period of 674 s.
As a result, the chance occurrence probability was
estimated as about 5\%,
which is consistent
with the evaluation using the normalized power spectrum.

The periodogram in panel (2-a) also reveals several weaker peaks,
including one at $\sim 340$ s
which can be identified with the 2nd harmonic in our power spectrum.
Although the fundamental,
which was rather weak in the power spectrum,
has become dominant in the periodogram,
the difference can be understood in the following way.
In the normalized spectrum in 
figure~\ref{fig:xmm-newton_pulse_search_fft} (right),
the power is expressed all relative to the local red-noise intensity,
so the fundamental becomes weaker due to the stronger local red noise.
In contrast, the fundamental peak becomes prominent in the periodogram,
because it sums over all the harmonics from $m=1$ (fundamental)
to $m=16$ (the Nyquist frequency for 32 bin),
whereas the 340 s peak is a sum over only $m=2, 4, ..,16$.
As a further attempt,
we detrended the first 8 ks of the XMM-Newton light curve,
using the 5th order polynomial.
However, no major changes took place
either in the power spectra or the periodograms.

\subsubsection{Folded pulse profiles}
%
For a further confirmation of the results derived so far,
we inspected the original light curve in
figure~\ref{fig:xmm-newton_lc},
and found that sharp intensity minima repeats at about 674 s,
as indicated by thin vertical lines over the
relevant portion of the data.
Therefore, the periodicity is considered to have good coherence.
In addition, such sharp intensity drops are
likely to arise via geometrical effects
(e.g., self eclipse of the emission region)
related to the rotation of the compact object
(a NS in this scenario).
This makes a contrast to the case of flares,
which would occur either at a specific rotation phase of the compact star,
or in a manner unrelated to the rotation
(e.g., caused by quasi-periodic blobs in the stellar winds).
Therefore, we regard the 674-s period
as a strong candidate for the pulse period of this X-ray source.
Its relation to the previously reported 346-s period
is considered in subsection 4.1.2.

\begin{figure}
\centering
  \includegraphics[width=120mm, angle=0]{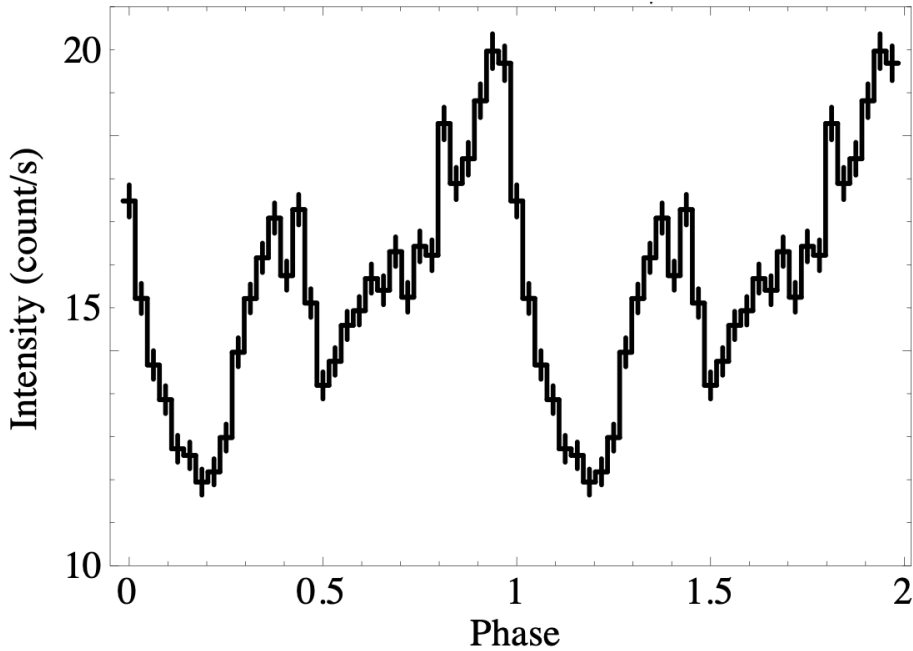}
 \caption{The 0.1--15 keV EPIC/pn data from the 1st 8 ks of the XMM-Newton observation, folded at 674 s.
  }
 \label{fig:xmm-newton_pulse_folded_lc}
\end{figure}

Figure \ref{fig:xmm-newton_pulse_folded_lc} 
shows the pulse profile,
obtained by folding the 0.1--15.0 keV data 
from the 1st 8 ks at the period of 674 s.

The pulse profile is rather structured,
in a qualitative agreement with the emergence of
up to the $m=8$ harmonics
in figure~\ref{fig:xmm-newton_pulse_search_fft} (right).
In addition, the profile is double-peaked,
consisting of a pair of large and small peaks
which are about half a cycle apart.
This explains why the even higher harmonics are strong
in the power spectra (either raw or normalized).

\subsection{Analysis of the brighter spectra}
To constrain the nature of the compact object of 
IGR J00370+6122 from X-ray spectral viewpoints,
we analyzed 1-10 keV energy spectra obtained
by the XMM-Newton/EPIC, the Suzaku/XIS, and the Swift/XRT.
Below, errors and upper limits of  the spectral parameters.
as well as those on fluxes and  luminosities,
all refer to 90\% confidence levels.

\begin{figure*}
\centering
 \includegraphics[width=170mm, angle=0]{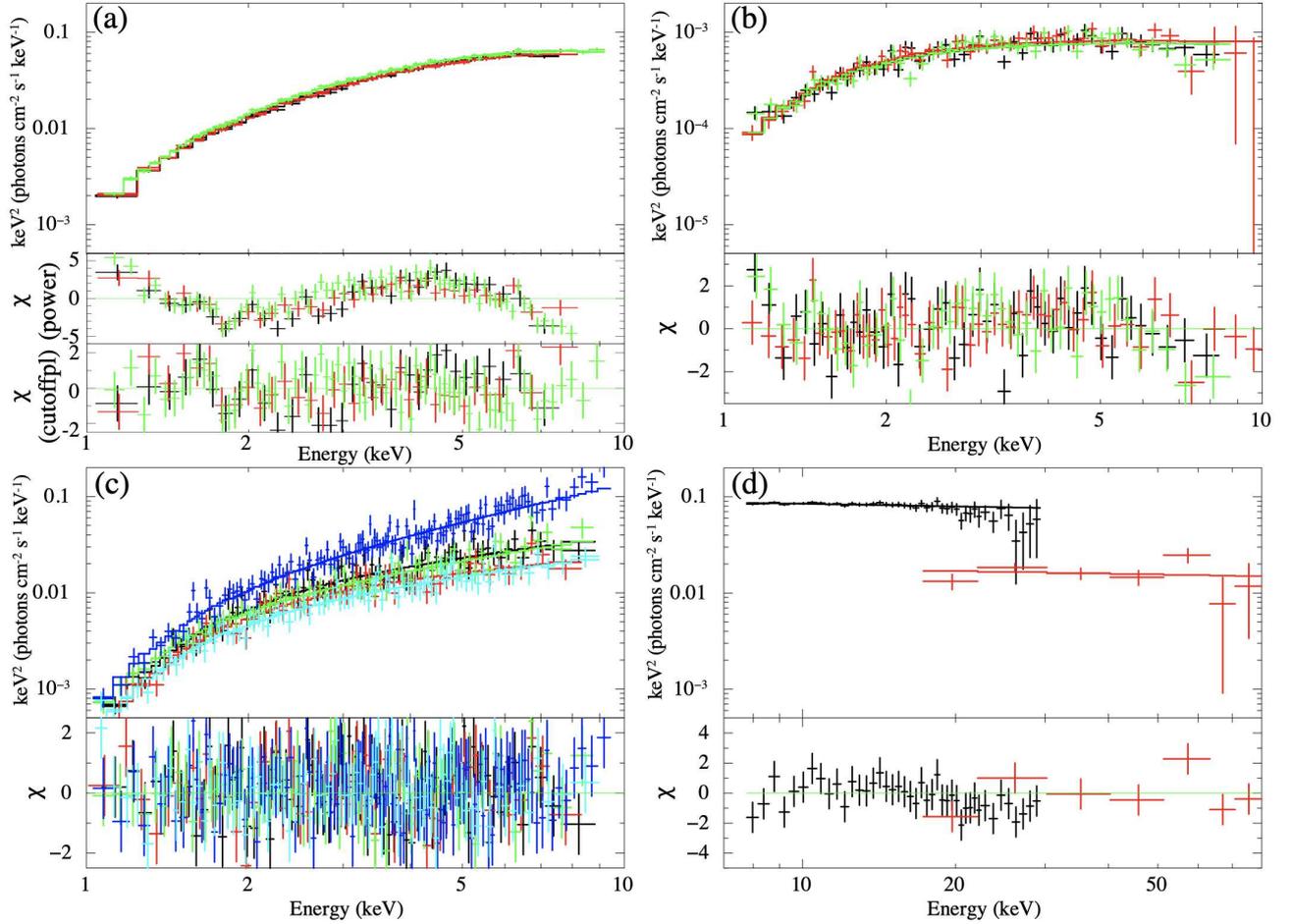}
 \caption{Model fitting results to the spectra of \IGR,
 presented in the $\nu F \nu$ form.
 (a) The EPIC MOS1 (black), MOS2 (red), and PN (green) spectra, 
 obtained in the flaring-state XMM-Newton observation,
 and  fitted simultaneously with an absorbed power-law model.
 The middle and bottom panels show residuals from the power-law fit
 and the cutoff power-law fit, respectively, both with absorption. 
 (b) The Suzaku XIS0 (black), XIS1(red), and XIS3 (green) spectra,
 fitted simultaneously with an absorbed power-law model.
 The fit residuals are given in the bottom panel.
 (c) The brightest 5 spectra from the Swift/XRT, fitted individually
 by an absorbed power-law model.
 (d) Combined spectra with the RXTE/PCA (black) and the INTEGRAL/ISGRI (red). They were obtained both near the periastron, but not simultaneous.
}
 \label{fig:All-spec}
\end{figure*}

We first analyze the PN/MOS1/MOS2 spectra,
acquired in the XMM-Newton observation in the flaring state 
(figure~\ref{fig:xmm-newton_lc}).
The data have the highest statistics,
and have allowed us to detect the 674-s tentative pulse period.
Because the spectra derived from the 1st 8 ks and the latter 15 ks 
do not show significant differences (except that in the normalization), 
below we employ the entire data.
As presented in figure \ref{fig:All-spec}(a) and table 2,
a simple power-law model with absorption 
({\tt phabs * powerlaw}) approximately reproduced the spectra,
incorporating a rather hard photon index of $\Gamma \sim 1.2$.
However, the fit was unacceptable due to 
residuals at $< 2$ keV and $>7$ keV.
When we employ a cutoff power-law model instead,
the residuals have disappeared and the fit has become acceptable (table 2).
The derived photon index is even harder, $\Gamma \sim 0$,
and the  cutoff energy is obtained as $E_{\rm cut} \sim 3.7$ keV.

The Suzaku/XIS data were obtained for 60 ks in quiescence,
when the source was about 40 times fainter 
than in the flaring-state XMM-Newton observation.
The source exhibited small (a factor of $< 4$) variability,
without significant pulsations.
As presented in figure \ref{fig:All-spec}(b) and table 2,
the time-averaged 1--10 keV XIS spectra 
were reproduced reasonably well by the simple absorbed power-law model,
with a softer photon index of $\Gamma \sim 2$.
When we apply the cutoff factor to the power-law continuum, the fit has been improved, with a decrease in chi-square by about 30,
and the cutoff energy has been constrained as $E_{\rm cut} \sim 3$ keV
(table 2 for details).

The Swift/XRT data sets have a typical exposure of a few kiloseconds each.
Although some of these data sets, 
particularly those when the source is active,
showed factor 3--10 intensity variations within each,
we again analyze the spectra which are averaged over individual observations.
Among the 31 data sets altogether,
nine (all acquired in flaring activity and near the periastron) 
have statistics which are high enough for the spectral studies.
As summarized in table 2, these 9 spectra have all been
reproduced successfully by the absorbed power-law model.
Figure \ref{fig:All-spec}(c) shows the brightest five spectra.

\begin{figure}
\centering
  \includegraphics[width=120mm, angle=0]{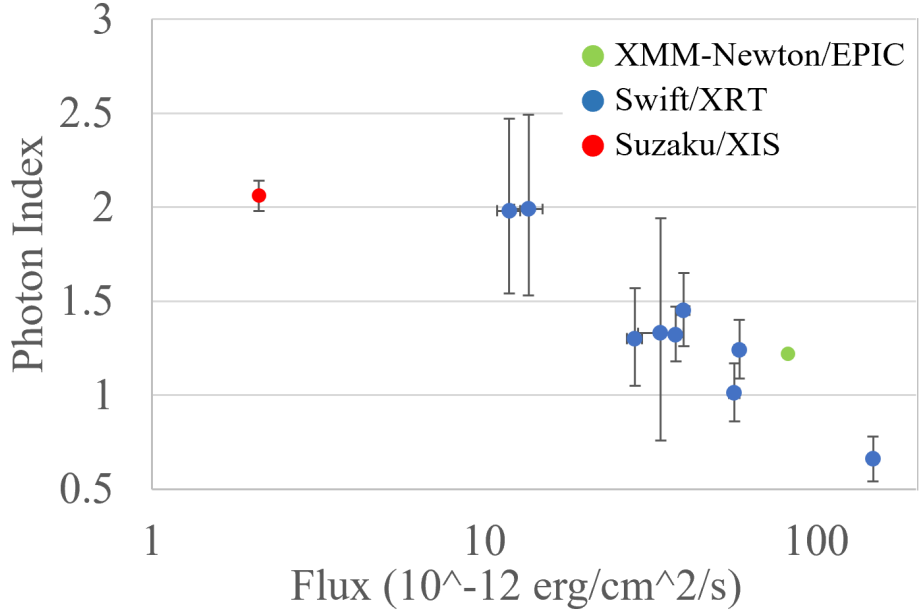}
 \caption{
 A Scattered plot between the photon index (from the power-law fit)
 and the 1--10 keV flux. 
 Data points refer to those presented in Table \ref{table:fitting_parameters},
 i.e., relatively bright data where the spectral shapes have been constrained.}
 \label{fig:flux_vs_photonindex}
\end{figure}

In figure \ref{fig:All-spec}(c),
we observe a spectral change with a “harder when brighter” trend;
the spectrum becomes harder ($\Gamma $ gets smaller)
when the source flux becomes higher.
For a further confirmation of this property,
figure \ref{fig:flux_vs_photonindex} shows a scatter plot
between $\Gamma$ and the 1--10 keV flux,
obtained with the XMM-Newton/EPIC, the Suzaku/XIS, 
and the Swift/XRT (the 9 data sets).
Thus, the “harder when brighter” property is clearly confirmed.

As in table 2, the absorbing column density $N_{\rm H}$ 
was relatively constant at about $ 1 \times 10^{22}$ cm$^{-2}$,
when the spectra from the different instruments
are fitted individually with a power-law continuum.
Since the high-statistics XMM-Newton spectra (ObsID=501450101) 
require a cutoff power-law continuum and a lower $N_{\rm H}$ value,
the actual absorption column density might be 
at around $0.7\times 10^{22}$ cm$^{-2}$.

\subsection{Analysis of the fainter spectra}

We have so far analyzed 11 spectra,
one from XMM-Newton, another from Suzaku, and 9 from the Swit/XRT. 
We still have one more XMM-Newton data set (the 2nd observation) 
when the source was very faint and was only in the MOS2 field-of-view,
and 22 Swift/XRT spectra acquired 
at orbital phases rather away from the periastron.
Although these data would not have statistics 
high enough to constrain the spectral properties,
they are still useful when we study 
how the flux depends on the orbital phase.
In dealing with these data,
we fitted simultaneously the source+background spectrum
and the background spectrum,
because some data bins would become negative
if we directly subtracted the background.
The source spectrum was modeled by a simple absorbed power-law, 
with $\Gamma=2.06$ and $N_{\rm H}=1.26 \times 10^{22}$ cm$^{-2}$ both fixed,
referring to the Suzaku/XIS determinations.
The background spectrum was derived 
from an outer region in the same detector,
and was modeled by an unabsorbed power-law,
with the photon index fixed at 
$\Gamma=0.20$ for the XMM-Newton/EPIC spectrum
and $\Gamma=0.64$ for those with the Swift/XRT.
In each spectral fitting, only the normalizations of 
the source and background components were set free.
The fit goodness was evaluated with the Poisson statics 
(C-stat in {\tt XSPEC}, \citealp{Cash79}),
instead of the chi-square criterion.

In the 2nd XMM-Newton observation (ObsID 0742800201), 
with a net exposure of 11 ks, 
we detected only 26 source+background events,
and 13 events in the background region.
From these, we obtained a 1--10 keV flux of 
$(8^{+5}_{-4}) \times 10^{-14}$ erg cm$^{-2}$ s$^{-1}$, 
which corresponds to a luminosity of 
$(1.1^{+0.7}_{-0.5}) \times 10^{33}$ erg s$^{-1}$ at 3.4 kpc.
Similarly, by analyzing the fainter 22 Swift/XRT spectra in this way,
we positively detected the source in 17 observations,
and derived upper limits in the remaining 5 cases.
In all of them, the 1--10 keV luminosity was $\lesssim 10^{34}$ erg s$^{-1}$. 
The derived luminosities (including the upper limits) are utilized later
in our examination of the orbital luminosity modulation.

\begin{table*}[htb]
\begin{center}
\tbl{Results of the spectral fitting, using the absorbed power-law and cutoff power-law models.}
{
  \begin{tabular}{lcccccc}\hline
    Instruments& ObsID & Flux$^*$& $N_{\rm{H}}$ ($10^{22}$ cm$^2$) & $\mit\Gamma$ &$E_{\rm{cutoff}}$ &$\chi^2$/d.o.f \\ \hline
    RXTE&
    91061010101$^{\dag}$ & 220 & 9.1$\pm$0.3 & 2.14$\pm$0.02 & -- &-- \\
    XMM-Newton&
    0501450101$^{\ddag}$  & 83.50 & 1.16 & 1.22 & -- & 1558.09/273 \\
    &0501450101$^{\S}$  & 79.82$^{+2.74}_{-2.64}$ & 0.62$^{+0.03}_{-0.03}$ & 0.05$^{+0.06}_{-0.06}$ & 3.70$^{+0.19}_{-0.17}$ & 316.79/272\\
    Suzaku&
    402064010   & 2.08$^{+0.28}_{-0.25}$ & 1.29$^{+0.11}_{-0.11}$ & 2.06$^{+0.08}_{-0.08}$ & -- & 193.25/147\\
    &402064010$^{\S}$   & 1.97$^{+0.53}_{-0.41}$ & 0.67$^{+0.21}_{-0.20}$ & 0.50$^{+0.47}_{-0.47}$ &  2.79$^{+1.19}_{-0.65}$& 161.62/146\\
    Swift&
    00032620005 & 58.70$^{+1.88}_{-2.00}$ & 1.35$^{+0.23}_{-0.21}$ & 1.24$^{+0.16}_{-0.15}$& -- & 84.32/78\\
    --&00032620009 & 39.92$^{+1.63}_{-1.69}$ & 1.45$^{+0.29}_{-0.26}$ & 1.45$^{+0.20}_{-0.19}$ & --&58.47/62\\
    &00032620013 & 11.92$^{+0.91}_{-0.95}$ & 2.09$^{+0.84}_{-0.73}$ & 1.98$^{+0.49}_{-0.44}$ & --&15.67/14\\
    &00032620015 & 13.62$^{+1.36}_{-1.28}$ & 2.90$^{+0.90}_{-0.79}$ & 1.99$^{+0.50}_{-0.46}$ & --&18.90/19\\
    &00032620019 & 56.75$^{+2.11}_{-2.20}$ & 0.82$^{+0.19}_{-0.17}$ & 1.01$^{+0.16}_{-0.15}$ & --&68.47/76\\
    &00032620028 & 34.04$^{+5.75}_{-4.97}$ & 1.88$^{+1.02}_{-0.89}$ & 1.33$^{+0.61}_{-0.57}$ & --&7.77/13\\
    &00032620029 & 28.47$^{+1.42}_{-1.50}$ & 1.96$^{+0.51}_{-0.45}$ & 1.30$^{+0.27}_{-0.25}$ & --&39.92/35\\
    &00032620030 & 148.48$^{+4.31}_{-4.47}$& 1.03$^{+0.18}_{-0.16}$ & 0.66$^{+0.12}_{-0.12}$ & --&114.42/129\\
    &00032620031 & 37.79$^{+1.15}_{-1.21}$ & 1.32$^{+0.20}_{-0.18}$ & 1.32$^{+0.15}_{-0.14}$ & --&58.35/88\\
    \hline  
    \end{tabular}}
    \label{table:fitting_parameters}
  \end{center}
\begin{tabnote}
\footnotemark[*] Unabsorbed fluxes in 1--10 keV
(3--20 keV for the RXTE/PCA data),
in units of $10^{-12}$ erg s$^{-1}$ cm$^{-2}$.\\
\footnotemark[$\dag$] Parameters refer to \cite{Zand07} for 3-20 keV. \\
\footnotemark[$\S$] Fitting results with the cutoff power-law model.
\end{tabnote}
\end{table*}

\subsection{High-energy spectral properties in bright phases}
To characterize the spectrum at higher energies 
when the source is near the periastron and is hence bright,
we analysed the RXTE/PCA (10--30 keV) and INTEGRAL/ISGRI (17.3-80 keV) data. 
The former was derived from standard data products 
for ObsID=91061-01-01-01,
which caught a flaring activity at an orbital phase of an orbital phase of $\sim$0.05 (or $\sim$0.14 in our phase definition in Table \ref{table:observation_parameter})
and enabled the detection of the 346-sec period (\citealp{Zand07}).
The utilized INTEGRAL/ISGRI data cover a 10-year period of 2003--2012,
but we selected the events only in the orbital phase interval from --0.1 to 0.1 where the source is significantly bright in figure \ref{fig:swift_bat_orbital_folded_lc}.
The RXTE/PCA and INTEGRAL/ISGRI spectra were fitted with 
a single power-law function without absorption.
The fit is simultaneous, except the spectral normalization 
which is allowed to take separate values between the two spectra
considering that they are not simultaneous.
As shown in figure \ref{fig:All-spec}(d),
the fit was successful with a reduced chi-square of 0.99,
and gave a flat continuum with a  photon index of $2.09 \pm 0.04$.
We also tried a cutoff power-law fit, 
in order to constrain a possible spectral turn-over 
which is observed in many other HMXBs.
However, the fit (which was already acceptable) 
did not improved significantly, 
and a 90\% confidence limit of $E_{\rm cut} >38.6$ keV was derived.
Thus, the 10--80 keV spectrum of \IGR\ can be explained
by a $\Gamma \sim 2.1$ power-law without significant high-energy cutoff.

\section{Discussion}

\subsection{The nature of the compact object in IGR J00370+6122}

In order to better understand the nature 
of the compact component in the HMXB IGR J00370+6122, 
we analyzed various X-ray archival data,
obtained with the 6 X-ray instruments onboard altogether 5 missions.
These data sets differ in many attributes such as the sensitivity,
energy range, observation epoch, time length, 
orbital-phase coverage,
and the source intensity during the observation.
Their comprehensive analysis has enabled us to 
deepen our understanding of this object.
Below, we summarize the obtained results
from the timing and spectral aspects.

\subsubsection{Timing results: orbital intensity variations}

Our analysis of the 15--50 keV Swift/BAT data, 
covering a 15 yr period of 2005--2019,
yielded the following three results.
First, the orbital period has been refined 
as in equation~(\ref{eq:Porb}).
Second, as already reported by \cite{Zand07},
the orbit-folded light curve in
figure~\ref{fig:swift_bat_orbital_folded_lc}
shows a strong orbital modulation,
that the X-ray emission is strongly enhanced
over a limited orbital phase from $-0.1$ to 0.2.
Finally, the same figure reveals a slight ($\sim 0.05$ cycle) delay
of the X-ray intensity maximum from the periastron passage,
in broad agreement with \cite{Grunhut14} and \cite{Gonzalez14}.

\begin{figure}
\centering
  \includegraphics[width=120mm, angle=0]{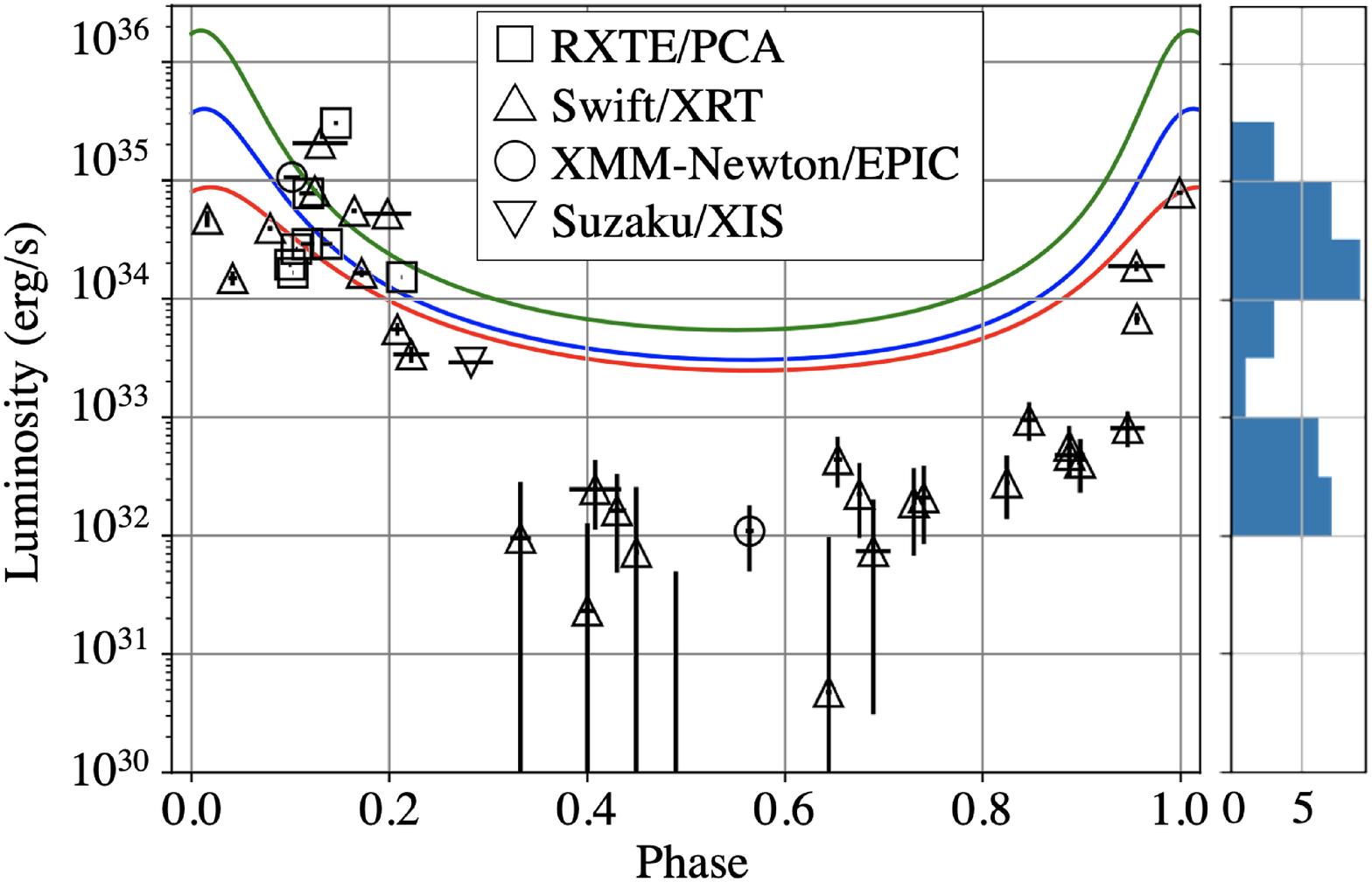}
 \caption{
 Orbital variations of the X-ray luminosity of \IGR.
 The data points of Swift/XRT, XMM-Newton/EPIC and Suzaku/XIS are from the present analyses, while RXTE/PCA ones are from \citealp{Zand07}.
 Three solid curves are theoretical predictions (see text),
 calculated for the stellar mass/eccentricity of 
 (red) $22~\MO/0.48$, (blue) $15~\MO/0.56$, 
 and (green) $8~\MO/0.56$, 
 referring to \cite{Grunhut14}, \citealp{Gonzalez14} 
 and \citealp{Hainich20}, respectively. 
 The right histogram shows the distribution of the measured luminosities
 excluding upper limits.}
 \label{fig:calculated_luminosity}
\end{figure}

Although the second result above is of importance,
the limited Swift/BAT sensitivity hampers us to infer 
from figure \ref{fig:swift_bat_orbital_folded_lc} 
to what extent the source gets dim when it is away from the periastron.
To understand this issue, in figure~\ref{fig:calculated_luminosity}
we plot all the flux measurements from the present study,
except those with the Swift/BAT,
as a function of the orbital phase.
Thus, away from the periastron phases,
the source has been found to decline to extreme low luminosities,
down to $\sim 10^{32}$ erg s$^{-1}$,
almost by 3 orders of magnitude from the periastron phases.
This behavior is likely to repeat regularly each orbital cycle,
because it is consistently indicated by the
relatively dense sampling with the Swift/XRT,
as well as by the Suzaku and the 2nd XMM-Newton observations.
In addition, we find a hint of a rather abrupt luminosity drop
at an orbital phase of $\sim 0.3$, 
and possibly a quick return at a phase of $\sim 0.95$.
Later, we try to theoretically explain this behavior.

\subsubsection{Timing results: the possible pulse detection}
To reconfirm the pulsation 
that was detected only once in the past by the RXTE/PCA,
we searched for a pulsation using the 2008 XMM-Newton data,
which are suitable for this purpose
because the object was rather bright and flare active
as in the RXTE observation,
and the data have a longer exposure (23 ks vs 4 ks).
As a result, a candidate period has been discovered at $674 \pm 15$ s.

\cite{Zand07} in fact found, in their RXTE periodogram, 
not only the 346 s peak which they favor,
but also a broad enhancement at about 700 s 
(which is higher than the 346 s peak) and 550 s.
Therefore, the two observations consistently prefer three common periods;
340-346 s, $\sim 550$ s, and 674-700.
Although the nature of the $\sim 550$ s enhancement is unclear,
the other two periods can most naturally be
regarded as the the fundamental and the 2nd harmonic,
as we already argued repeatedly.
If the folded pulse profile is mildly time variable,
the major and minor peaks in figure \ref{fig:xmm-newton_pulse_folded_lc}
could sometimes have comparable intensities,
to make the pulse period appear to be $\sim 340$ s.
In fact, the X-ray pulsar Vela X-1 was first 
thought to have a pulse period of about 141 s,
but the period was later revised as 283 s \citep{Mcclintock76}.

\subsubsection{Spectral results}
Next, the energy spectral analyses were performed
to characterize \IGR\ from spectral viewpoints.
The 1-10 keV energy spectra obtained by the XMM-Newton/EPIC, 
the Suzaku/XIS, and the Swift/XRT were approximately 
represented by an absorbed power-law model.
Furthermore, a mild spectral cutoff with $E_{\rm cutoff} =3-4$ keV,
together with a very hard spectral slope with $\Gamma=0-0.5$,
are indicated by the XMM-Newton spectrum,
and by the Suzaku data to a lesser extent.
The absorption column density was stable at 
$N_{\rm H} \sim 1 \times 10^{22}$ cm$^{-2}$.
Since the optical companion, BD+60 73, 
has an optical extinction of $A_{\rm v}=2.39$ \citep{Gonzalez14, Hainich20},
the Galactic line-of-sight column is estimated as 
$N_{\rm H} \sim (5-6)\times 10^{21}$ cm$^{-2}$,
which is consistent with the Galactic coordinates of \IGR, 
$(\ell, b)=(121^\circ.2, -1^\circ.5)$, and its distance.
Thus, the observed $N_{\rm H}$ is likely to be a mixture 
of about equal Galactic and circum-source contributions.
Analyzing the 3--20 keV RXTE/PCA spectrum with an absorbed power-law model,
\cite{Zand07} derived $N_{\rm{H}}\sim 10^{23}$ cm$^{-2}$ (table \ref{table:fitting_parameters}), 
which is an order of magnitude higher than those we have obtained.
This discrepancy probably arose 
because the intrinsic spectral curvature was mimicked by the strong absorption,
when the RXTE/PCA spectrum above 3 keV was fitted with a single power-law.

\begin{figure}
\centering
  \includegraphics[width=120mm, angle=0]{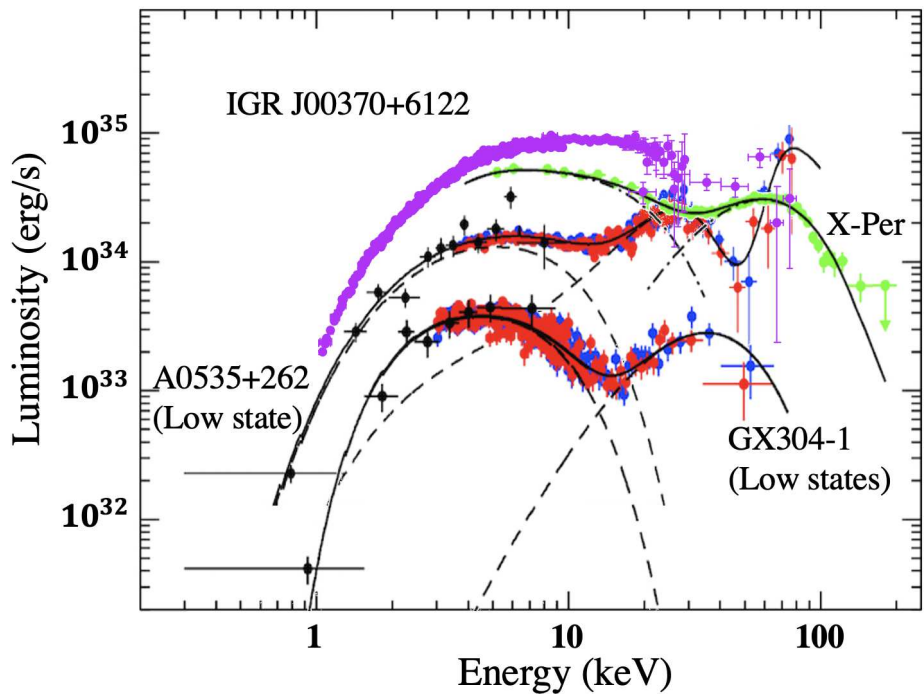}
 \caption{A combined spectrum with XMM-Newton/EPIC, RXTE/PCA 
 and INTEGRAL/ISGRI of IGR J00370+6122 is compared with those of
 X Persei, A0535+26 and GX304-1 adapted from \citet{Droshenko12},
 \citet{Sergey19a} and \citet{Sergey19b}. 
 For IGR J00370+6122, the XMM-Newton/MOS2, PN, RXTE/PCA and INTEGRAL/ISGRI spectra 
 are renormalized in reference to the XMM-Newton/MOS1 spectrum.}
 \label{fig:spc_comparison}
\end{figure}

\subsubsection{Identification of the compact object in \IGR}

Even though the evidence for the 674-s pulsation gives
a strong support to the argument for a magnetized NS in this system,
we are still left with a few percent probability 
that the detection is false.
In addition, the pulsation was detected only from a limited portion
of the XMM-Newton observation.
Similarly, the evidence for pulsation had been obtained 
only once, out of a number of past observations.
Therefore, we need to carefully examine whether our interpretation
can be supported from other pieces of evidence.

Starting our examination from a basic viewpoint,
the compact object in \IGR\ must be either a white dwarf,
a black hole (BH), or a NS.
The maximum luminosity of $\sim 10^{35}$ erg s$^{-1}$
and the 674-s pulsation (assuming it to be real)
are both explicable in terms of a white dwarf.
However, it would be extremely unlikely 
that a white dwarf ever formed 
a binary with a massive star. 
The observed spectrum, with weak or no emission lines, 
also disagrees with line-rich thermal spectra from accreting white dwarfs.
Therefore, the white dwarf scenario can be readily ruled out.

Let us next consider a possibility that
the compact star is a BH. 
In fact, the observed violent variability
and the relatively hard spectra both
broadly agree with those of accreting BHs.
If so, however, the luminosity of this object,
$10^{35}$ erg s$^{-1}$ when it is bright,
would translate to 0.02\% of the Eddington luminosity
for a typical BH mass of $5~\MO$.
Then, the BH would have to be in the low/hard state,
wherein the spectrum from a few keV to a few tens keV 
should be approximated by a $\Gamma \sim 1.7$ power law,
accompanied by a clear cutoff with $E_{\rm cutoff}=50-150$ keV
(e.g., \citealp{Makishima08}).
The observed spectrum of \IGR\ is clearly distinct,
harder than this in $<10$ keV, 
and softer but possibly less curved in $>10$ keV.
Thus, the compact object cannot be considered as a BH, either.

As a result, we are left with the sole option
that the compact object is a  NS.
In this case, we still have to tell whether the NS is strongly 
magnetized with the MFs $B\gtrsim 10^{12}$ G,
or has relatively weak MFs as $B\lesssim 10^{9}$ G.
Clearly, the former is favored for a few reasons.
First, as an old stellar population,
weak-field NSs form binaries predominantly
with low-mass stars, and rarely with massive ones.
(Probably \IGR\ is located on the Perseus arm.)
Second, the former type of objects often exhibit
flare-like variations like those seen in \IGR\
on time scales of minutes,
whereas the latter objects, when dim, vary randomly 
on times scales of minutes to hours, but more continuously,
typically with a relative amplitude of $\sim 20\%$ 
(e.g., \citealp{Takahashi11}).
Finally, the spectrum of \IGR\ is very similar
to those of binary X-ray pulsars (see next subsection),
whereas distinct from those of weak-MF NSs 
(e.g., \citealp{Ono17})
which are more similar to those of BHs in the low/hard state.
Therefore, independently of the 674-s periodicity,
our comprehensive study supports 
the presence of a high-MF NS in \IGR.

\subsection{Comparison with the known binary X-ray pulsars}
\label{subsection:comparizon}

Taking it for granted 
that the compact object in \IGR\ is a magnetized NS, 
our next task is to compare it with the known binary X-ray pulsars, 
particularly those with low luminosities,
to understand its properties and behavior
as a mass-accreting NS.

\subsubsection{General similarities}

We have successfully 
quantified the 1--10 keV spectra of \IGR\,
over a rather wide luminosity range 
as seen in figure \ref{fig:All-spec},
and found that the photon index becomes harder from 2 to $<1$
as the luminosity increases from $3\times10^{33}$ 
to $2\times10^{35}$ erg s$^{-1}$ (figure~\ref{fig:flux_vs_photonindex}).
This spectral behavior, “harder when brighter”, 
and the very hard photon index of $< 1$, 
agree with the properties observed from nine Be-type X-ray pulsars 
when they are $< 0.1$ times the Eddington luminosity (Reig \& Nespoli \citeyear{Reig13}).
This similarity 
provides yet another supporting evidence
that \IGR\ is an accreting magnetized NS.
The “harder when brighter” trend, consistently observed
from these low-luminosity pulsars (including \IGR),
may be explained as a spectral hardening caused by 
an increase of the inverse-Compton scattering probability,
in response to the increase in the accretion rate \citep{Ralf17}.

We find several additional similarities between \IGR\ and Be-type pulsars.
These include the large orbital intensity changes
(figure~\ref{fig:calculated_luminosity}),
the small but significant difference between the peak-brightness 
and periastron phases (figure~\ref{fig:swift_bat_orbital_folded_lc}),
and the weakness of iron-K emission line (figure~\ref{fig:All-spec}).
Considering further the lack of a circumstellar disk 
\citep{Reig05} in the primary BD+60 73, 
its stellar winds may be somewhat anisotropic, 
and denser along equatorial directions of its rotation.

As seen so far,
the magnetized-NS interpretation is supported,
not only via the elimination argument,
but also by general similarities seen
between \IGR\ and the known binary X-ray pulsars
(particularly of Be companions).

\subsubsection{Magnetic fields and spin periods}

Using the RXTE/PCA and INTEGRAL/ISGRI data (though not simultaneous),
we have found that the high-energy spectrum of \IGR\ extends
up to 80 keV with a photon index of $\Gamma \sim 2.2$,
without noticeable cutoff.
Combined with the information below 10 keV from XMM-Newton, Suzku, 
and Swift, the X-ray spectrum of \IGR\ is hence found to keep
a rather flat shape over nearly two orders of magnitude in energy. 
Actually, the constraint of $E_{\rm cutoff} >38.6$ keV derived 
for \IGR\ is considerably higher 
than the values of $E_{\rm cutoff}=5-10$ keV measured from 
binary X-ray pulsars at a typical luminosity of $\sim 10^{37}$ erg s$^{-1}$.
This difference may be attributed
to the very low luminosity of \IGR,
$\sim 1 \times 10^{35}$ erg s$^{-1}$, 
or  0.1\% of the Eddington luminosity 
for a NS with 1.4 $M_{\odot}$,
even though it refers to the brightest phase near the periastron.

To examine the above possibility, 
we compare in figure \ref{fig:spc_comparison}
the broad-band spectrum of \IGR\ 
with those of three binary X-ray pulsars with comparable low luminosities,
all having Be-type optical companions.
One is the nearby binary X-ray pulsar X Persei,
which has a persistently low luminosity 
(\citealp{Droshenko12, Sergey19a, Sergey19b, Yatabe18}).
The other two are typical recurrent Be-type transients, 
A0535+26 and GX $304-1$,
which were recently observed in a very X-ray dim phase
during a decay phase of their outbursts \citep{Ralf17, Sergey19b}.
Thus, the most outstanding spectral property 
common to the four objects is the hard X-ray continua
that extend with a flatter shape to higher energies,
than those of more luminous accreting pulsars
which strongly turn over at $\gtrsim 20$ keV.
Therefore, the flat spectrum of \IGR\ may be explained,
at least partially, 
as a property which is common to dim  X-ray pulsars,
even though its mechanism is unclear at present.

Another noticeable property in figure \ref{fig:spc_comparison},
which is particularly prominent in A0535+26 and GX $304-1$,
is a two-hump structure of the spectrum, 
with two peaks at a few keV and a few tens keV,
as already pointed out by \cite{Sergey19a}  and \cite{Sergey19b}.
(In A0535+26, another sharp dip at 50 keV is
due to the CRSF; \citealp{Terada07}).
This two-hump structure is possibly present 
in X Persei and in \IGR\ as well, but less visible. 
Among the three comparison objects,
A0530+26 and GX $304-1$ have surface MFs of 
$B \sim 4.5 \times 10^{12}$ G and $\sim 5.5 \times 10^{12}$ G,
respectively, as measured securely with their CRSFs.
In contrast, 
X Persei has a considerably stronger dipole MF of $B \sim$10$^{14}$ G,
as estimated through the GL79 modeling 
(Introduction; \citealp{Yatabe18})
which has been verified and calibrated quantitatively using 
long-term MAXI observations 
\citep{Sugizaki15, Takagi16, Sugizaki17, Sugizaki20}.
Then, the closer resemblance of the \IGR\ spectrum
to that of X Persei,
than to those of the two recurrent transients,
suggest that \IGR\ also has a MF of $> 10^{13}$ G, 
as already suggested by \citet{Grunhut14}.
This inference is also supported by an empirical positive dependence
of $E_{\rm cutoff}$ on the CRSF energy \citep{Makishima99};
this dependence may be explained theoretically
in terms of cyclotron resonance cooling in the accretion column 
which will shift to higher energies for stronger MFs.

Interestingly, the three comparison objects in figure \ref{fig:spc_comparison} 
have relatively long spin periods on the order of 100 s or longer.
Other examples of low-luminosity pulsars 
with  hard flat spectra and very long periods 
include 4U~0114+65 (with a period of 9340 s), 
4U~1954+317 ($\sim 2\times 10^4$ s), and 4U~2206+54 (5540 s).
In contrast, the Be pulsars with fast rotation,
4U~0115+63 (3.6 s) and X0331+53 (V0332+53; 4.4 s), 
show rather softer spectra when they are dim 
at $\lesssim 10^{34}$ erg s$^{-1}$ (Wijnands \& Degenaar \citeyear{Wijnands16}).
Therefore,
an X-ray pulsar with a low luminosity and a hard continuum spectrum
tends to have a long spin period.
This empirical tendency
suggests \IGR\ to have a long spin period as well,
and reinforces the case for the 674-s pulsation.

\subsection{The luminosity change along the orbit and magnetic propeller effects}

As discussed in section 4.1.1, we have found that the luminosity of 
\IGR\ changes along the orbit by a large dynamic range reaching
three orders of magnitude (figure \ref{fig:calculated_luminosity}).
We further notice an abrupt luminosity drop at the orbital phase $\sim 0.3$,
from $\sim 4 \times10^{33}$ to $\sim 1\times10^{32}$ erg s$^{-1}$,
and possibly a reversal at the orbital phase $\sim 0.95$. 
Let us numerically examine whether this behavior can be 
explained by a standard scenario of the wind-capture accretion.

Presuming that the NS in \IGR\ is powered by 
capturing the stellar winds from the massive primary,
we employ a simple Hoyle-Lyttleton accretion picture,
taking into account the orbital phase dependence of the wind capture rate
(Wind-Rose effect, Natalya \& Lipunov \citeyear{Natalya98}). 
In this picture,
the mass accretion rate $\dot{M}_{\rm acc}$ onto the NS
is assumed as
\begin{equation}
    \dot{M}_{\rm acc} = \frac{1}{4}\left(\frac{R_{\rm cap}}{r}\right)^2
   \left( \frac{v_{\rm rel}}{v_{\rm w}} \right) 
    \dot{M}_{\rm w}~.
\end{equation}
Here,
$r$ is the distance of the NS from the primary star,
$\dot{M}_{\rm w}$ is the primary's wind mass loss rate, 
$v_{\rm rel}$ is the relative velocity 
between the NS star and the local stellar wind,
$v_{\rm w}$ is the stellar-wind velocity at the position of the NS, and $R_{\rm cap}$ is the gravitational wind-capture radius defined as
\begin{equation}
R_{\rm cap} = \frac{2 G M_{\rm NS}}{v_{\rm rel}^2}~,
\end{equation}
where 
$G$ is the gravitational constant, 
and 
$M_{\rm NS}$ is
the NS mass.
In the calculation, 
we assume the stellar wind from the primary to be isotropic, 
and utilize the wind velocities numerically given in table A.7 of 
\cite{Hainich20} as a function of the distance from the star,
which is assumed to have a radius of 16.5 $\RO$.
We adopt $\dot{M}_{\rm acc}=3\times 10^{-8}~\MO$ yr$^{-1}$
from \cite{Hainich20}.
The orbital  motion of the NS is calculated using 
the orbital period of equation (\ref{eq:Porb}), 
together with three cases for the primary's mass 
and the orbital eccentricity as described in the caption.
The X-ray conversion factor is set to be unity,
i.e., the gravitational energy gained via accretion 
is all converted to the radiation, 
assuming a NS radius of 12 km and a NS mass of 1.4 $\MO$.
Any other complex effects (e.g., a photoionization of the stellar wind by the X-ray) are not taking into account in this calculation.

The X-ray luminosity changes calculated in this way are superposed
in figure~\ref{fig:calculated_luminosity} on the observed data,
where the phase 0 is aligned with the periastron passage time of equation (2).
The model curves have the peak at an orbital phase of 0.02--0.03,
This is because $v_{\rm rel}$
becomes minimum right after the periastron passage. 
In contrast, the observed luminosity peak is 
more delayed from the periastron epoch.
As already mentioned in subsection \ref{subsection:comparizon},
this discrepancy may be attributed to some 
anisotropy of the stellar wind properties.
We do not discuss this issue any further.

The calculated orbital modulation of the luminosity is
thus 1.5 to 2.5 orders of magnitude,
which arises mainly due to changes in $v_{\rm rel}$,
hence those in $R_{\rm cap}$,
along the eccentric orbit.
These are significantly smaller than the observed dynamic range 
(3 orders of magnitude).
Even when a systematic normalization adjustment
(due, e.g., to an estimation error of $\dot{M}_{\rm w}$)
is allowed between the observation and calculation,
this discrepancy is too large to be attributed to 
systematic errors of any of the assumed parameters.
For example, to explain the observation,
the orbital eccentricity would have to be extremely large as $>0.9$,
in contradiction to the optical observations.
Alternatively, the stellar winds should have to be accelerated
to have a twice higher velocity at the apastron 
than at the periastron; this is also unlikely.
A still more important point is 
that none of these ideas would be able to explain the 
markedly dim phase interval observed between the orbital phases of $\sim$ 0.3 and $\sim$ 0.95.
To explain this phenomenon, 
we need to invoke some additional effects,
in addition to the simple Hoyle-Lyttleton accretion scenario.

One possibility is the X-ray photoionization of the stellar winds,
which may cause large luminosity changes along the orbit \citep{Bozzo20}.
To evaluate this effect, we calculated 
the ionization parameter $\xi$ at the periastron.
Among the three conditions considered in the present work,
the shortest distance between the primary's surface
and the NS is $r \sim 10~\RO$,
where the stellar-wind density becomes 
$n \sim 10^{11}$  cm$^{-3}$ \citep{Hainich20}.
The maximum X-ray luminosity is $L \sim 10^{36}$ erg s$^{-1}$.
These conditions yield $\xi\equiv L/n r^2\sim 20$ erg s$^{-1}$ cm$^{-1}$.
Therefore, the photoionization would not work 
sufficiently
in \IGR\,
because it is considered inefficient even in Vela X-1,
which has a much higher value of $\xi \sim 10^4$ \citep{Bozzo20}.
Even if the photoionization ever worked, 
it would produce two sharp peaks
in the orbital X-ray light curve \citep{Bozzo20},
one preceding and the other following the periastron,
with a relatively reduced luminosity in between.
The observed behavior of \IGR\ is opposite,
showing an abrupt luminosity increase over an orbital phase
near the periastron
between $\sim$0.3 and 0.95.
Therefore, we conclude that the photoionization effect
cannot explain the light curve of this system,
and that other explanations must be sought for.

The observed long pulse period and the low X-ray luminosity
make \IGR\ very reminiscent of X-Persei,
which has $B\sim 10^{14}$ G (see section 1).
Similarly, comparing three fundamental radii of accreting NS binaries, 
i.e. accretion, corotation, and Alfv\'en radii, \cite{Grunhut14} argued
that \IGR~ possibly has a strong magnetic field of $\sim 10^{15}$ G.
If so, we naturally expect the operation of  so-called propeller effect,
which has actually been observed in some NS binaries
(e.g., \citealp{Asai13, Tsygankov16}).
In short, the gravitationally captured material can 
accrete onto the NS (accretion mode)
when the matter density is high and hence the Alfv\'en radius 
is smaller than the co-rotation radius.
When the matter density decreases to below a certain threshold,
the pulsar magnetosphere will expand beyond the co-rotation radius,
to prevent the direct mass accretion onto the NS.
The latter  is often called ``propeller mode'',
because the rotating magnetosphere would expel the matter off the NS.
This picture can explain the very low luminosity 
away from the periastron as a continued propeller-mode phase,
and the observed luminosity jumps as the epochs of 
transitions between the accretion and propeller modes.
This inference is strengthened by the histograms given
on the right side of figure~\ref{fig:calculated_luminosity},
which indicate that the luminosity takes a bimodal distribution,
with a gap between $\sim 10^{33}$ and $\sim 10^{34}$ erg s$^{-1}$.

Let us apply the propeller effect modeling to the case of \IGR.
If the transition between the accretion mode and the propeller mode
takes place at an X-ray luminosity $L_{\rm X}$,
where the gravitational pull and the magnetic pressure are balanced,
the dipolar MF strength $B$ (expressed by its value on the surface)
is given as
\begin{eqnarray}
B 
 &\sim 5.7\times10^{13} \zeta^{-\frac{7}{4}} 
\left(\frac{L_{\rm X}}{10^{33} \rm{ \ erg \  s^{-1}}}\right)^{\frac{1}{2}} \left(\frac{P}{10^3\;\rm{s}}\right)^{\frac{7}{6}} 
\quad \rm{G}
\label{eq:torque_equilibrium}
\end{eqnarray}
where $P$ is the spin period 
and $\zeta \sim$1 is a numerical factor (\citealp{Makishima16}).
Employing $P=674$ s and the mode transition luminosity as
$\sim 2\times10^{33}$ erg s$^{-1}$ from figure \ref{fig:calculated_luminosity},
we obtain $B \sim 5\times 10^{13}$ G.
As clear from the above equation, this high value of $B$
is a direct consequence of the long spin period.
In addition, the very low luminosity of \IGR\ can be explained
naturally  as a consequence of magnetic inhibition
of the wind capture process \citep{Yatabe18}.
Thus, IGR J00370+6122, like X Persei, 
is thought to have a dipole MF 
which is considerably higher than those of typical binary X-ray pulsars,
and is even comparable to those of magnetars.
This inference agrees with the suggestion derived in subsection 4.2
based on our broadband spectral comparison.
In this case, the CRSF should appear around 600 keV,
which would be hard to detect.
This is consistent with an apparent lack of the CRSF
in the 1--80 keV spectrum of figure \ref{fig:spc_comparison}.

We have so far invoked the GL79 torque equilibrium theory
and a strong MF to explain the low luminosity and the slow spin period.
The application of the GL79 scheme is supported
by a series of successful calibrations (see Introduction),
and the case of the two Be-type pulsars,
4U~0115+63 (3.6 s pulse period) and X0331+35 (V0332+53, 4.4 s).
In a declining phase of their outbursts,
these objects showed a clear luminosity drop,
to be interpreted as due to the propeller effect
\citep{Tsygankov16},
at luminosities of $1.4 \times 10^{36}$ erg s$^{-1}$
and $2.0 \times 10^{36}$ erg s$^{-1}$, respectively.
Then, equation~(\ref{eq:torque_equilibrium}) gives
$B=2.5 \times 10^{12} \zeta^{-\frac{7}{4}}$ G for 4U~0115+63
and $B=2.5 \times 10^{12} \zeta^{-\frac{7}{4}}$ G for X0331+53,
which agree very well with their surface MF strength
measured with the CRSF, 
$1.2 \times 10^{12}$ G and $3.0 \times 10^{12}$ G,
respectively \citep{Makishima16}.
Of course, some reservations may be needed,
because the presence of a stable accretion disk, assumed in the GL79 scheme, 
may not necessarily be applicable to the present object.
If, instead, so-called quasi-spherical accretion (QSA) 
\citep{Shakura12, Shakura14} takes place,
the required MF strength would be reduced to $\sim 10^{12}$ G
(e.g., \citealp{Sanjurjo17}).
For example, the SFXT IGR J11215-5952 is 
characterized by a persistently low luminosity, 
a long spin period, and periodic flares at periastron 
(\citealp{Sidoli20} and references therein), 
and is very reminiscent of \IGR.
\citet{Sidoli20} tried to explain its behavior
using the QSA framework.
Considering however the richer observational verification
of the GL79 scheme than that of the QSA,
and the independent evidence from the harder and flatter spectral continuum,
we would still favor the strong-MF interpretation of IGR J00370+6122.

This conclusion will be strengthened 
if follow-up observations of \IGR\ in the future 
confirms the rapid luminosity jump
at the orbital phases $\sim 0.3$ and $\sim 0.95$.

\section{Conclusion}
By analyzing archival X-ray data of \IGR\ obtained with XMM-Newton,
Suzaku, Swift, RXTE, and INTEGRAL, we have obtained the following results.
\begin{enumerate}
\item The orbital period has been refined. 
Near the periastron, the source brightens up to 
$\sim 2 \times 10^{35}$ erg s$^{-1}$,
and often exhibits flaring activity.
The luminosity peak is delayed by $\sim 0.05$ orbital cycles 
from the periastron passage.
\item From the 1st 8 ks of the 2008 XMM-Newton observation made near periastron, 
a possible periodicity was detected at 674 s.
This strongly suggests that the compact object is an accreting magnetized NS
rotating rather slowly at this period.
Because the pulse profile folded at 674 s is double peaked,
the previously reported 346 s period could be half 
(or twice in frequency)
that observed in the present work.
\item Through the analysis of 1--10 keV spectra obtained with 3 instruments,
the “harder when brighter” trend has been confirmed.
This makes \IGR\ similar to known binary pulsars accreting at low luminosities.
\item A combined 10--80 keV spectrum can be described by a 
single power-law with $\Gamma \sim 2.2$ without noticeable cutoff. 
Its similarity to the spectrum of X Persei empirically suggests
that \IGR\ has a considerably higher MF than ordinary X-ray pulsars.
\item Away from the periastron,
the luminosity has been found to decrease by 3 orders of magnitude, 
down to $\sim 1 \times 10^{32}$ erg s$^{-1}$,
involving abrupt luminosity jumps.
These can be explained by the magnetic propeller effect,
and a MF of $5\times 10^{13}$ G is derived.
\end{enumerate}

\begin{ack}
This work was supported by JSPS KAKENHI Grant Number 19J13685.
\end{ack}

\bibliography{mybibliography}

\end{document}